\documentclass{desyproc}
\usepackage{multicol}

\makeatletter

\newenvironment{figurehere}
{\def\@captype{figure}}
{}
\makeatother

\begin{document}

\def\e{\rm e}
\def\GeV{{\rm Ge\!V}}


\title{A new look at Multiple Parton Collisions}



%
%
%
%
%
%
%
%
%

\author{{\slshape Yuri L.\ Dokshitzer} \\[1ex]
LPTHE, Paris-VI, France and PNPI, Gatchina, Russia}


\acronym{MPI@LHC 2011}
\maketitle
\def\beq{\begin{equation}}
\def\eeq{\end{equation}}


\begin{abstract}
The key ingredients of systematic QCD analysis of MPI are discussed.  
\end{abstract}



\section{Introduction}

This contribution presents the QCD approach to studying MPI that I am involved in together with 
Boris Blok, Lonya Frankfurt and Mark Strikman. 
The basic new ideas of our approach have been announced in a short article \cite{BDFS1}. 
A detailed pQCD analysis of the main contributions to double hard parton collisions, 
as well as evolution of the emerging generalized double parton distributions, $_2$GPD's, can be found in \cite{BDFS2}.  

An excess of {\em jets + photon}\/ events in the back-to-back kinematics have signaled the presence of double parton collisions   
in the Tevatron experiments \cite{Tevatron1,Tevatron2,Tevatron3}. This phenomenon cannot be explained within 
a naive independent parton approximation. A model of two-proton correlations inside the proton that is capable 
of explaining the magnitude of the MPI contribution will be presented in \cite{BDFS3}. 

Our approach to the problem of MPI is in certain sense opposite to that developed by Tevatron experiments.  
The CDF group in the pioneering study \cite{Tevatron1} has formulated a noble quest of 
extracting the MPI contribution without referring to either QCD theory or even to MC event generators.  
Such a puristic approach has successfully performed a noble task of establishing the presence double hard collisions. 
However, it does not help much in developing the {\em theory}\/ of the phenomenon that would stem from 
the first QCD principles at the ``hard'' end, and exploit known phenomenology of hadron interactions at the ``soft'' end. 
Meantime, such a theory is necessary for quantifying two-parton correlations in the proton and for better understanding
of the underlying physics of collider experiments.

Following the pioneering work of Refs.~\cite{TreleaniPaver,mufti},
a number of theoretical papers on multiparton interactions appeared in recent years \cite{Treleani,Wiedemann,Frankfurt,Frankfurt1,SST}.  
They were based on the parton model and geometrical picture of collisions in the impact parameter space.
This topic is being intensively discussed in view of the LHC program \cite{Perugia,Fano};
Monte Carlo  event generators that produce multiple parton collisions are being developed \cite{Pythia,Herwig,Lund}.
In our view, however, important elements of QCD that are necessary for theoretical understanding
of the multiple hard interactions issue have not been properly taken into account by above-mentioned intuitive approaches.

More recently, theoretical papers exploring the nature and properties of double parton distributions 
and discussing their QCD evolution have appeared \cite{BDFS1,Diehl,DiehlSchafer,Ryskin}.

The problem of theoretical approach to MPI is, sort of, educational: both the probabilistic picture, the MC generator technology is based upon,
and even the Feynman diagram technique, when used in the momentum space, prove to be inadequate
for careful analysis and understanding of the physics of multiple collisions.


\section{Hidden reefs of MPI analysis}

A careful approach to MPI phenomena uncovers a number of unconventional features.   

Thus, in order to be able to trace the relative distance between the partons, one has to use the mixed longitudinal
momentum--impact parameter representation which, in the momentum language, reduces to introduction
of a mismatch between the transverse momentum of the parton in the amplitude and that of the same parton
in the amplitude conjugated.

Another unusual feature of the multiple collision analysis may look confusing at the first sight. It is the fact that,
even at the tree level, {\em the amplitude}\/ describing the double hard interaction contains additional integrations
over longitudinal momentum components; more precisely --- over the difference of the (large) light-cone momentum components
of the two partons originating from the same incident hadron (see Section~\ref{SubSec:PT34}).

\subsection{Transverse structure}
Cross section of two-parton collision can be cast in the following intuitively clear form:
\beq\label{eq:1}
d\sigma^{(h)} = \int d^2\rho_1 \!\! \int d^2\rho_2\> f(x_1,\rho_1) f(x_2,\rho_2)\cdot d\sigma^{(p)}; \quad  B=\rho_2-\rho_1. 
\eeq
Here vectors $\rho$ are the transverse positions of incident hadrons 1 and 2 with respect to the point where the two partons interact;  
their difference is the impact parameter of the hadron collision $B$. 
Local parton density $f$ is the square of the wave function:
$ f(x_1,\rho_1) \>=\> \psi(x_1,\rho_1) \, \psi^\dagger (x_1,\rho_1)$,
or, in the momentum representation, 
\beq\label{eq:3}
f(x_1,\rho_1) = \int \frac{d^2k_{1\perp}}{(2\pi^2)} \psi(x_1,k_{1\perp})  \int \frac{d^2k'_{1\perp}}{(2\pi^2)} \psi^\dagger(x_1,k'_{1\perp}) \cdot \e^{i\rho_1(k_{1\perp}-k'_{1\perp})}.
\eeq
Substituting 
into \eqref{eq:1} and integrating over transverse coordinates gives
\beq\label{eq:41}
\int \frac{d^2k_\perp}{(2\pi)^2}\, \psi(x, k_\perp) \int \frac{d^2k'_\perp}{(2\pi)^2}\, \psi^\dagger(x, k'_\perp) \times \int d^2\rho\> 
\e^{i\rho(k_\perp - k'_\perp)} = \int  \frac{d^2k_\perp}{(2\pi)^2}\> \psi(x, k_\perp) \psi^\dagger(x, k_\perp).
\eeq
\begin{multicols}{2}
This shows that one could have written an answer for the cross section in terms of momenta from the start, 
treating incident objects as plane waves with given (longitudinal and transverse) momentum components. 

The situation is different when two pairs of partons collide. 
Indeed, in this case the transverse coordinates of four partons are not independent but are related, see Fig.~\ref{Fig1}:
\[
   \rho_1-\rho_3 = \rho_2-\rho_4 = B.
\]

\begin{figurehere}
\centering
{\includegraphics[width=160pt]{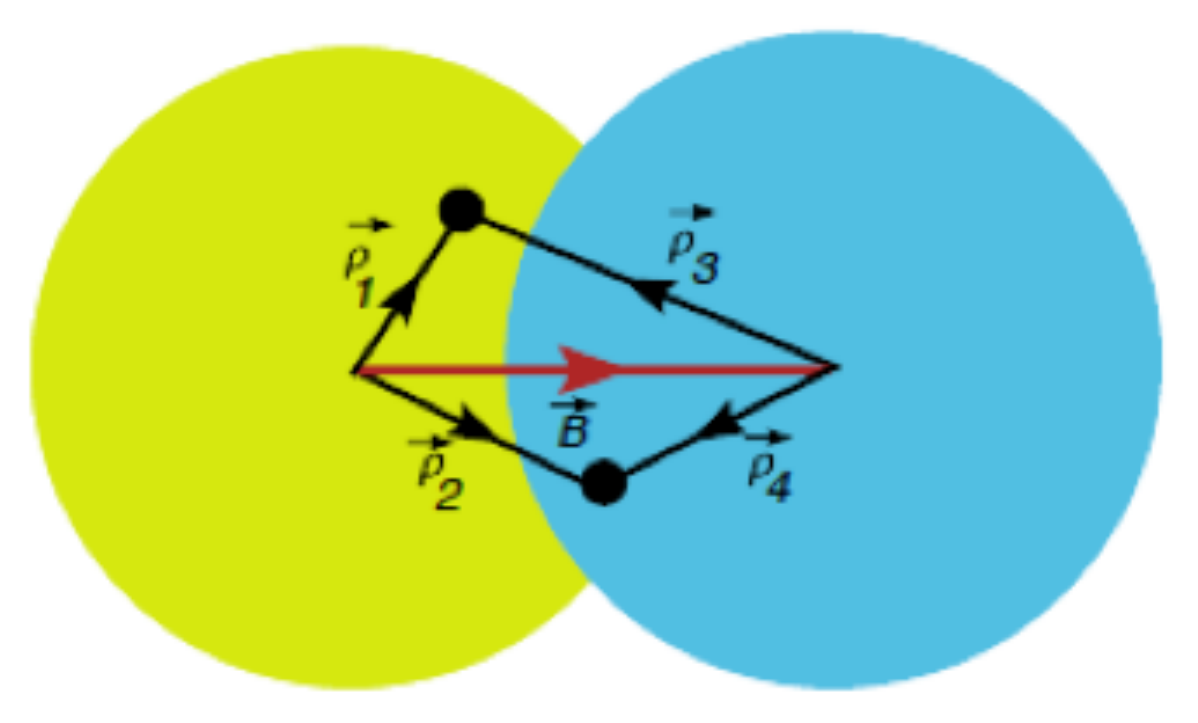}}
\caption{\label{Fig1}Geometry of four-parton collision}
\end{figurehere}
\end{multicols}

To see how does this condition affect the momentum picture, we introduce inclusive two-parton probability density distribution,
and turn to the momentum representation in analogy with Eq.~\eqref{eq:3}.

\beq\label{eq:42}
\begin{split}
D(x_1,x_2;\rho_1,\rho_2) = & \sum_{n=3}^\infty \int \prod_{k=3}^n \bigl[d^2\rho_k dx_k\bigr] \cdot \delta(\sum_{i=1}^n x_i\rho_i) \\
&\cdot \psi(x_1,\rho_1;x_2,\rho_2; x_3,\rho_3;\ldots x_n,\rho_n)\,
 \psi^\dagger (x_1,\rho_1;x_2,\rho_2; x_3,\rho_3; \ldots x_n,\rho_n), 
\end{split}
\eeq
Then, integrations over $(\rho_1+\rho_2)$, $(\rho_3+\rho_4)$ and $(\rho_1-\rho_2)+(\rho_3-\rho_4)$ produce, respectively,
\beq
k_{1\perp}- k'_{1\perp} = - (k_{2\perp}- k'_{2\perp}) \, \equiv \Delta, \quad 
k_{3\perp}- k'_{3\perp} = - (k_{4\perp}- k'_{4\perp}) \, \equiv \tilde{\Delta} \quad \mbox{and} \>\>  \tilde{\Delta} = -\Delta.
\eeq

\begin{multicols}{2}
The presence of the relation 
\[
   \rho_1-\rho_3 - (\rho_2-\rho_4)=0
\] 
leaves the transverse momentum parameter $\Delta$ arbitrary. 

We conclude that, in the language of momenta, a mismatch appears between transverse momenta of the parton in the wave function 
and the wave function conjugated. This mismatch is the same for all four participating partons as shown in Fig.~\ref{Fig2}.

Thus, the new variable $\Delta$ is an intrinsic part of the two-parton correlation function  in 
\begin{figurehere}
\centering
 \resizebox{\columnwidth}{!}{ \includegraphics{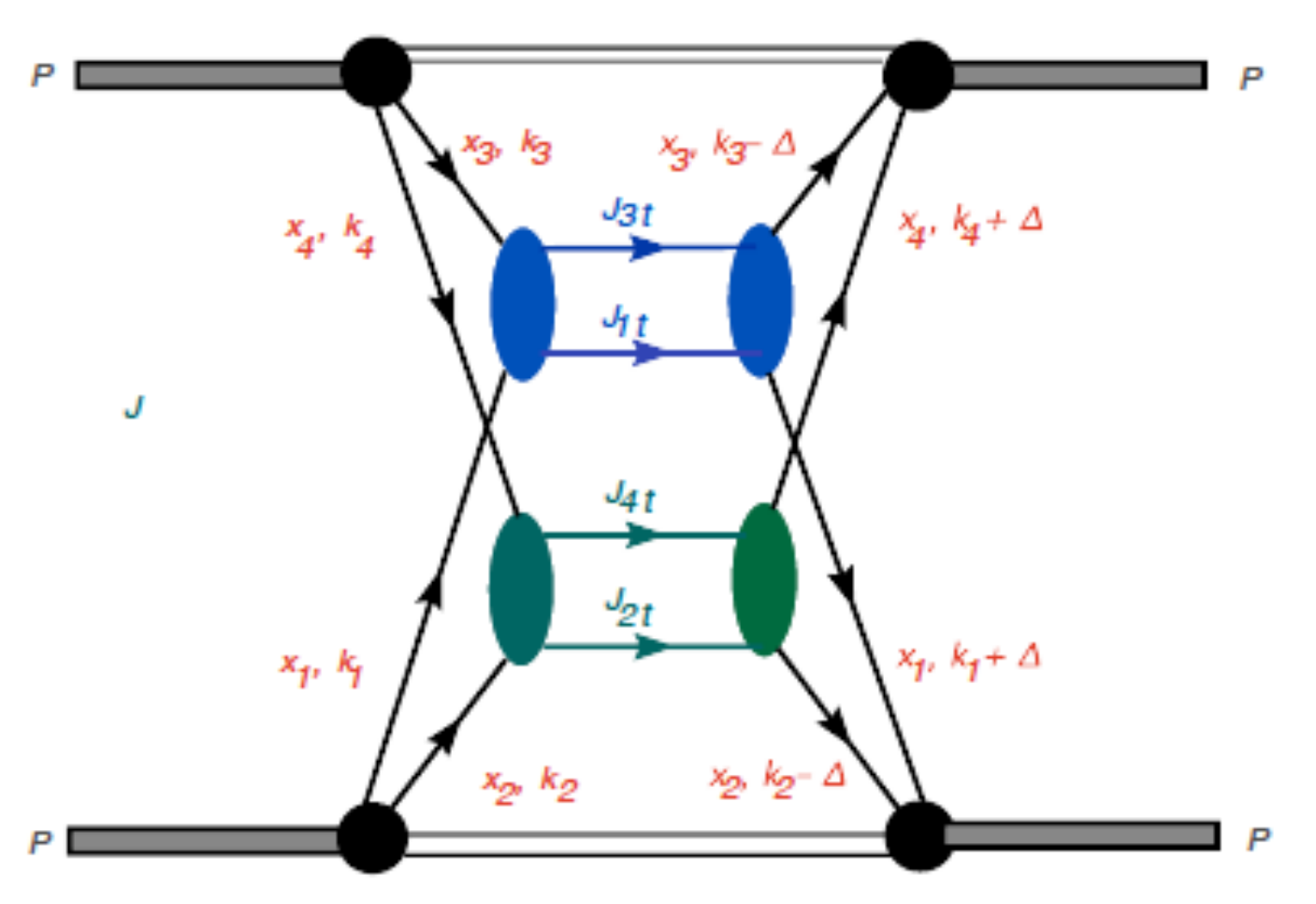}}
\caption{\label{Fig2} Shifts in parton momenta }  
\end{figurehere}
\end{multicols}
\vspace{-4mm}
\noindent
momentum space, that was dubbed in \cite{BDFS1} ``the generalized double parton distribution'', $_2$GPD:  

\beq
_{[2]}D_h^{ab}(x_1, x_2; q_{1}^2, q_{2}^2; \Delta).
\eeq 
Here $a,b$ mark parton species, $h$ --- the hadron, and $q_1^2,\, q_2^2$ (within the usual logic of parton distributions) stand for the corresponding ``hardness scales'': the upper limits of (logarithmic) integrations over parton transverse momenta, $k_{1\perp}^2$ and $k_{2\perp}^2$.
This is what concerns the transverse space structure.

\subsection{Longitudinal structure \label{SubSec:PT34}}
\begin{multicols}{2}
Now we shall look at {\em longitudinal momenta}\/ of participating partons.

To elucidate the problem that one encounters here it is instructive to examine the case when a parton ``0'' from one hadron virtually splits into two, ``1'', ``2'', which offspring partons enter two hard interactions with partons ``3'' and ``4'' from the second hadron. 

This situation is shown in Fig.~\ref{Fig3}, where black blobs mark two hard interactions that 

\begin{figurehere}
\centering
 { \includegraphics[width=75pt]{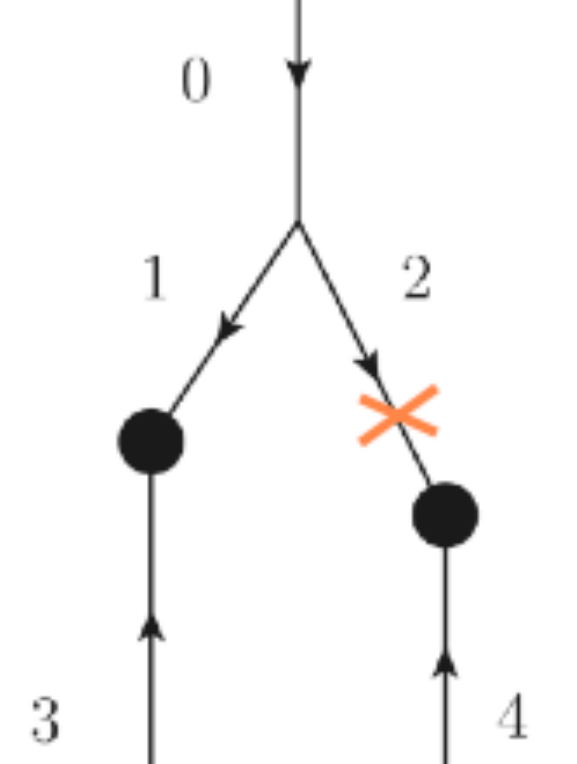}}
\caption{\label{Fig3}On-mass-shell singularity} 
\end{figurehere}
\end{multicols}
\noindent
produce some large mass final state systems
 (intermediate bosons, pairs of large transverse momentum jets, etc.). 

Fig.~\ref{Fig3} is a tree amplitude. 
This means that knowing the momenta of incident partons 0, 3, 4, and the 4-momenta of the produced final state systems
$Q_{13}$ and $Q_{24}$, one unambiguously determines  the momenta of the virtual state partons 1 and 2.  

The problem is, at certain values of longitudinal momenta of incident partons 
one of the intermediate partons can go {\em on mass shell}. For example, the parton line ``2'' in Fig.~\ref{Fig3}.
The amplitude develops a strange singularity right inside the physical region of the external momenta 
($k_0^2, k_3^2, k_4^2 \le0$, $Q_{13}^2$ and $Q_{24}^2$ positive).   

This singularity has been discussed in the literature more than once (see, e.g., \cite{stirling,stirling1}). 
However, to the best of my knowledge, 
its meaning and significance remained unclear before the explanation that we gave in \cite{BDFS1}.

Let us go back to the usual hard process picture, e.g., to DIS scattering. 
Consider perturbative splitting of an incident parton ``0'' into ``1'' and ``2'', of which the former experiences 
hard scattering (gets hit by a lepton with a large momentum transfer $Q^2$), while the latter goes into the final state. 

\begin{figure}[h]
\begin{center}
\includegraphics[width=200pt]{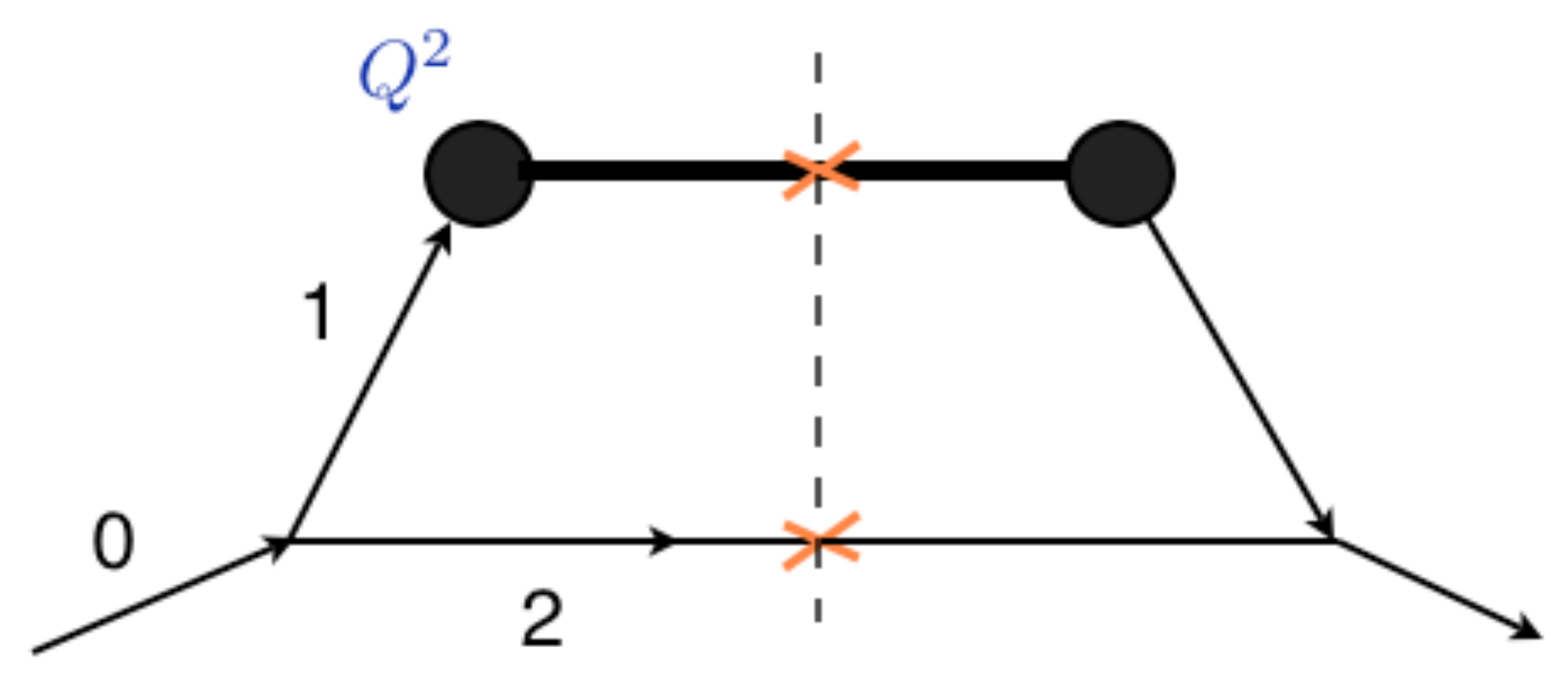}
\end{center}
\caption{\label{Fig4}``On-mass-shell'' partons in deep inelastic scattering}
\end{figure}

When calculating the DIS cross section (see Fig.~\ref{Fig4}) we put the parton ``2'' on mass shell and do not trace its fate. 
It may split developing a final state jet, it may propagate as a ``real particle'' at macroscopically large distances 
(confinement does not concern us here) and might eventually enter another hard interaction. 
This is exactly what happens with the diagram of Fig.~\ref{Fig3} and where lies an explanation of the origin of that disturbing singularity.    

What happens is the following. The singularity appears at definite momenta of incident particles and, in particular, of partons ``3'' and ``4''.
Definite momenta mean plain waves. But plain waves are not localized in space--time, so that the distance between the two hard interactions in Fig.~\ref{Fig3} is not known and can be in fact arbitrarily large. 
If the hard scattering of ``1'' and ``3'' would occur in the LHC tunnel, and the collision of ``2'' and ``4'' --- in, say, Gran Sasso, then 
the presence of the singularity is natural: it would correspond to free propagation of the particle ``2'' between Geneva and L'Aquila.  
Such a scenario is possible, but this is not what we are looking for: we intend instead to study  
the situation when ``3'' and ``4'' belong to the same proton! 

To assure spatial localization inside one hadron, one has to construct a wave packet by smearing over the relative longitudinal momentum
of the two partons. 
Importantly, the kinematics of the process determines only the {\em sum}\/ of the light-cone momentum components ($\beta_3+\beta_4$). 
So one is allowed --- and has to --- introduce an integration over the {\em difference}\/ of the two momenta, $\beta_3-\beta_4$, at the amplitude level. 
This smearing eliminates the singularity of the diagram Fig.~\ref{Fig3}: the integral reduces to the residue at the pole of the propagator ``2''.

\section{Generalized double parton distributions }
We have chosen to study production of two pairs of large transverse momentum jets as an example of double hard interaction. 
The corresponding cross section is conveniently represented as a product of cross sections of two independent collisions normalized 
by the factor $S$ that has dimension of area: 
\beq\label{eq:Dint}
 \frac{d\sigma^{(4)}}{dt_1 dt_2} = \frac{d\sigma(x_1,x_2)}{dt_1} \,\frac{d\sigma(x_3,x_4)}{dt_2} \times \frac1S,
   \quad  \frac{1}{S} = \frac{\int\frac{d^2\Delta} {(2\pi)^2}\, D(x_1,x_2;\Delta)D(x_3,x_4;-\Delta)}{D(x_1)D(x_2)D(x_3)D(x_4)}.
\eeq
The quantity $S$ is often referred to in the literature as ``effective cross section''. 
However, a cross section, by definition, depends on {\em interaction strength}\/ 
while $S$ characterizes transverse area of two-parton correlation in a hadron and longitudinal correlation between the partons (see \cite{BDFS3})
\beq
\frac{d\sigma^{(4\to4)}}{dt_1 dt_2} \propto  \left(\frac{\alpha_s^2}{Q^4}\right)^2 \cdot R^{-2} \>=\> \frac{\alpha_s^4}{R^2\,Q^8}
=  \frac{\alpha_s^4}{Q^6}\cdot  \frac{1}{R^2Q^2} \>\propto\>  \frac{d\sigma^{(2\to4)}}{dt_1 dt_2} \cdot \frac{1}{R^2Q^2}.
\eeq
So, this cross section turns out to be {\em power suppressed}\/ as compared with that of the $2\to4$ jet production mechanism when  
one gets {\em four jets}\/ out of 2-parton collision at an expense of two additional QCD emissions, $d\sigma^{(2\to4)}$.

\subsection{4- and 3-parton collisons}
The value of the correlation radius is determined by convergence properties of the $\Delta$ integral in \eqref{eq:Dint}.
If two partons are taken directly from the non-perturbative hadron wave function, $_2$GPD=$_{[2]}D(x_1,x_2)$, 
the correlation area is of the order of the transverse size of the hadron. 

There is an additional contribution to the MPI cross section due to collision of {\em three}\/ partons. 
In this case the numerator of \eqref{eq:Dint} has a mixed structure:
\[ 
_{[1]}D(x_1,x_2;\Delta)_{[2]}D(x_3,x_4;-\Delta) \>+\>  {}_{[2]}D(x_1,x_2;\Delta)_{[1]}D(x_3,x_4;-\Delta) .
\]
Here $_{[1]}D$ stands for 2-parton distribution involving {\em perturbative}\/ parton splitting, as in the upper part of Fig~\ref{Fig3}. 
Being a small-distance correlation, $_{[1]}D$ depends on $\Delta$ only logarithmically (via parton evolution effects). 
As a result, the $\Delta$ integration in \eqref{eq:Dint} becomes broader. In spite of this {\em geometrical enhancement}, 
the $3\to4$ contribution turns out to be numerically small at Tevatron energies ($x>10^{-3}$) but may become significant at the LHC \cite{BDFS3}. 

Evolution equations that incorporate into $_2$GPDs  all-order radiative QCD effects in the leading collinear approximation are described in \cite{BDFS2}.

\subsection{A four-parton or a two-parton collision?}

There were discussions in the literature whether the process of double parton splitting shown 
in Fig.~\ref{Fig24} should be looked upon as a 4-parton collision. 

On one hand, it looks indeed as two hard interactions of four partons. On the other hand, such a diagram naturally appears as a loop correction to a ``normal'' $2\to4$ QCD process when one goes beyond the tree approximation. The question of potential {\em double counting}\/ was raised.

\begin{multicols}{2}
\parbox{180pt}{
\begin{figurehere}
\centering
 { \includegraphics[width=90pt]{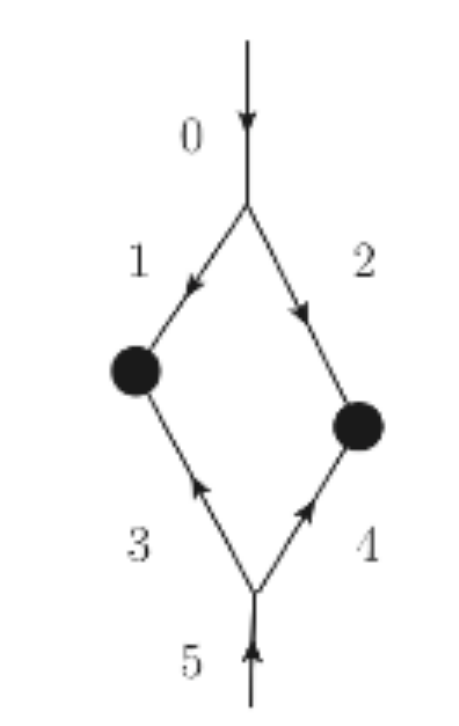}}
\caption{\label{Fig24} Loop in a double hard process} 
\end{figurehere}}

The process displayed in Fig.~\ref{Fig24} is a ``product'' of two small-distance correlations, $_{[1]}D(x_1,x_2)_{[1]}D(x_3,x_4)$. 
Since $_{[1]}D$ practically does not depend on $\Delta$, the integral in \eqref{eq:Dint} formally diverges. 

This means that this double hard interaction is not a MPI, in our interpretation (\cite{BDFS1,BDFS2}). 
It lacks a characteristic feature of MPI, namely a power enhancement of the differential cross section 
in the back-to-back kinematics, $Q_{13}^2, Q_{24}^2 \ll Q^2$ (see below).  

This is a loop correction that belongs to $2\to4$ background and has to be subtracted in a search for MPI.
\end{multicols}
\noindent

\subsection{Modeling $_2$GPD}
The first natural step is an {\em approximation of independent partons}, which allows one to relate $_2$GPD with known objects, namely
\beq
D(x_1,x_2, q_1^2,q_2^2;\Delta) \simeq G(x_1,q_1^2;\Delta) G(x_2, q_2^2;\Delta).
\eeq
Here $G$ is the non-forward parton correlator (known as generalized parton distribution, GPD) 
that determines, e.g., hard vector meson production at HERA (Fig.~\ref{Fig5}). 

\begin{multicols}{2}
\begin{figurehere}
\centering
\includegraphics[width=140pt]{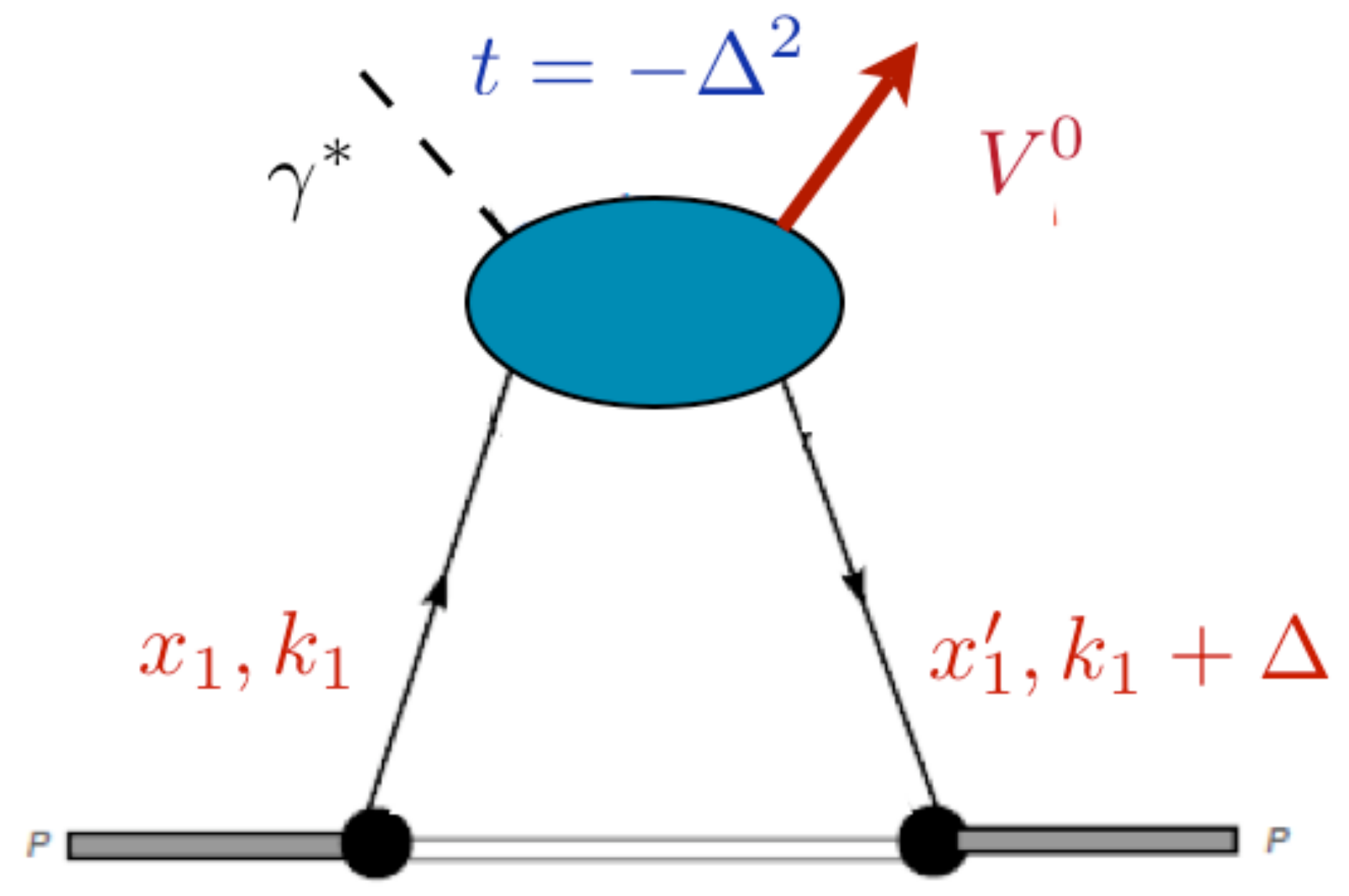}
\caption{\label{Fig5} GPD in VM electroproduction}
\end{figurehere}

The GPD, on its turn, can be modeled as 
\beq
 G(x_1,q_1^2;\Delta) \>\simeq\> D(x_1,q_1^2) \times F_{2g}(\Delta^2) ,
\eeq
with $D$ --- usual one-parton distribution determining DIS structure functions and $F$ --- the two-gluon form factor of the hadron. 

The latter is a non-perturbative object; it falls fast with the ``momentum transfer'' $\Delta^2$. 
\end{multicols}
\noindent
This form factor can be parametrized differently. For example, by a dipole formula:
\beq
     F_{2g}(\Delta^2) = \left( 1+ \frac{\Delta^2}{m_g^2}\right)^{-2}.
\eeq 
Here $m_g^2$ is an effective parameter whose value extracted from HERA data lies in the ballpark of 
$m_g^2(x\sim 0.03, Q^2\sim 3\,\GeV^2)\simeq 1.1\,  \GeV^2$.

A simplistic approximation of independent partons does not answer the call: it fails to explain a factor 2 enhancement 
of back-to-back 4-jet production observed by Tevatron experiments \cite{Tevatron1,Tevatron2}. 
So, intra-hadron correlations between partons have to be taken into account. 
One does not know much about them a priori. However, certain information about non-perturbative 2-parton correlations can be extracted from 
phenomenology of {\em inelastic diffraction}, which allows one to construct a viable model for the $_2$GPD of a nucleon \cite{BDFS3}.

\section{Differential distribution in back-to-back kinematics}
Four-parton interaction is a ``higher twist'' eventuality. 
The fact that the total MPI cross section is  {\em power suppressed}\/ as compared with the $2\to4$ cross section does not mean that $4\to4$ and $3\to4$ collisions are impossible to access at high $Q^2$.   

There is an essential difference between the two 4-jet production mechanisms. 
Namely, in $2\to4$ processes the final jets form a ``{\em hedgehog}\/'', while 
double parton collisions, $4\to4$, produce two pairs of nearly {\em back-to-back jets}.  
Actually, in the {\em back-to-back kinematics}\/ the two channels become comparable. 
 
\begin{figure}[h]
\begin{center}
\includegraphics[width=250pt]{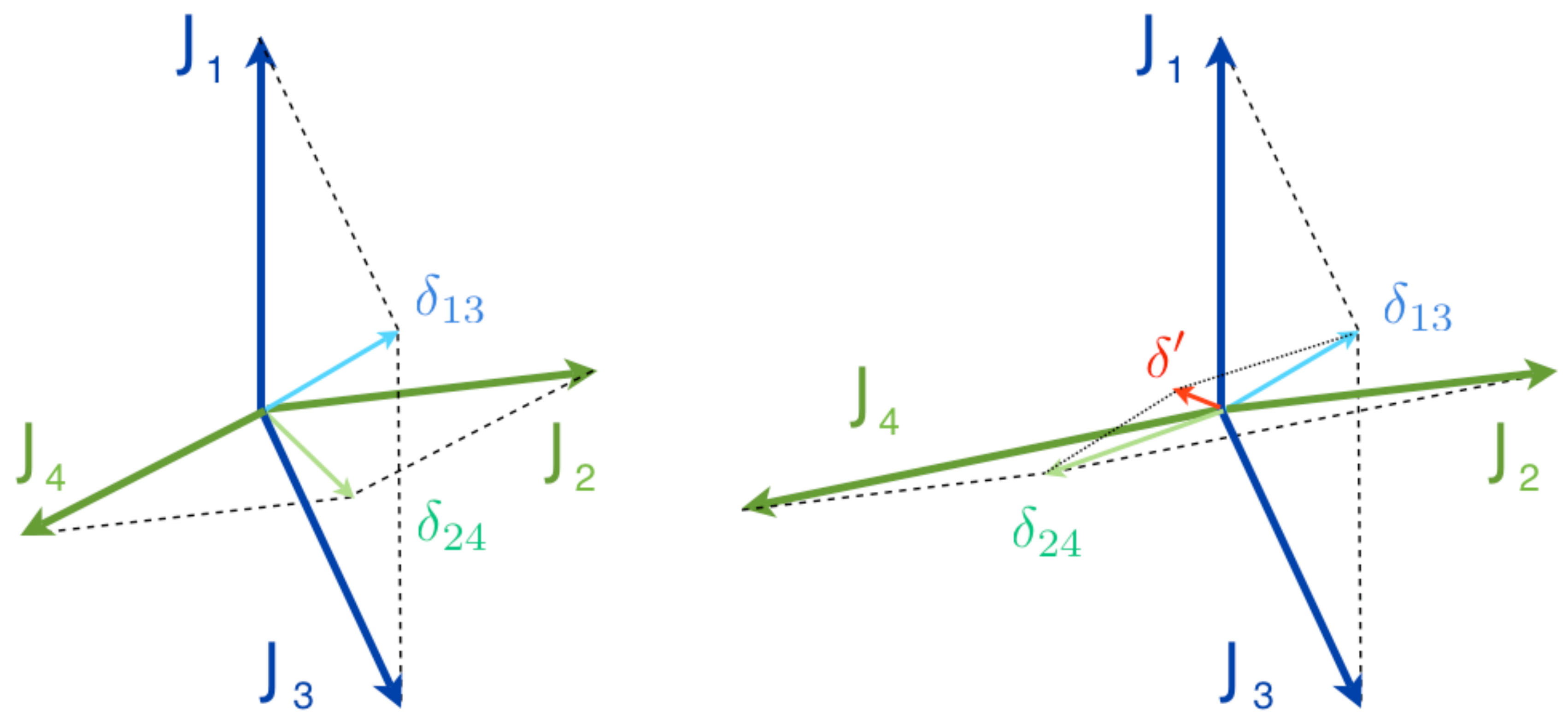}
\end{center}
\caption{\label{Fig6} Enhanced kinematical configurations of four jets}
\end{figure}

Differential distributions due to $4\to4$ and $3\to4$ processes exhibit {\em double collinear enhancement}\/: they peak at small {\em jet imbalances}, 
$\delta_{ik}^2\ll p_{i\perp}^2\simeq p_{k\perp}^2$, Fig.~\ref{Fig6}. 
\begin{subequations}\label{eq:diffs}
\begin{eqnarray}\label{eq:diffs1}
\left. \frac{d\sigma}{dt_1dt_2\, d^2\delta_{13} d^2\delta_{24}} \right/  \frac{d\sigma}{dt_1dt_2} 
&\propto& \frac{\alpha_s^2}{\delta_{13}^2\,\delta_{24}^2}, \quad \delta_{ik} = p_{i\perp}+p_{k\perp}; \\
\label{eq:diffs2}
\left. \frac{d\sigma}{dt_1dt_2\, d^2\delta_{13} d^2\delta_{24}}  \right/  \frac{d\sigma}{dt_1dt_2} 
&\propto&  \frac{\alpha_s^2}{\delta'^2\,\delta^2}, \quad
\delta^2=  \delta_{13}^2   \simeq \delta_{24}^2 \gg (\delta_{13}+\delta_{24})^2 \equiv  \delta'^2.
\end{eqnarray}
\end{subequations}
Structure of singularities displayed in \eqref{eq:diffs1} --- independent enhancements in two pair imbalances --- is typical for $4\to4$ processes.   

\begin{figure}[h]
\begin{center}
\includegraphics[width=250pt]{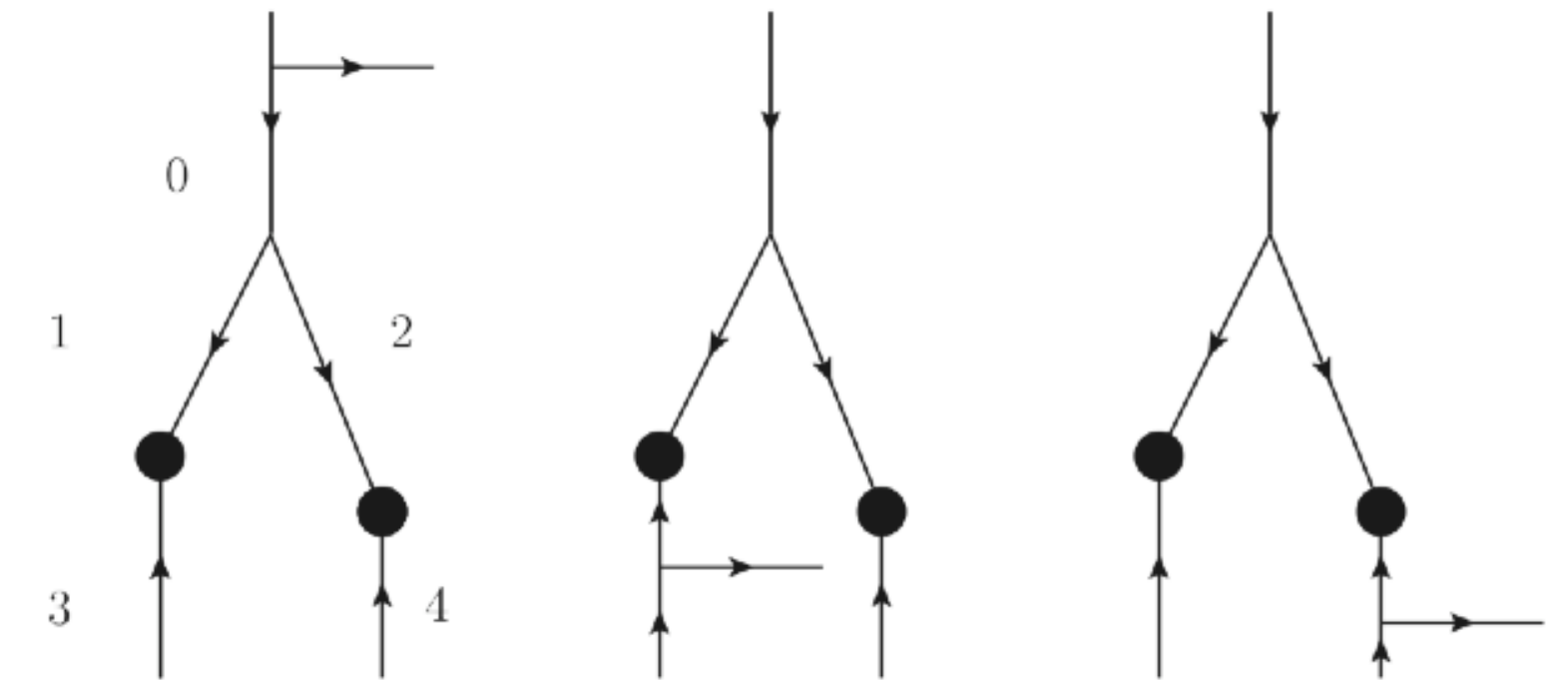}
\end{center}
\caption{\label{Fig8} Origin of $\delta'^2$ singularity in Eq.~\eqref{eq:diffs2}}
\end{figure}
The ``end-point'' contributions due to $3\to4$ configurations with no QCD emissions between the parton splitting ($0\to 1+2$ in Fig.~\ref{Fig8}) 
and the two hard vertices is enhanced as \eqref{eq:diffs2}. 

Singularities in Eqs.~\eqref{eq:diffs} get smeared by double logarithmic Sudakov form factors of the partons involved, 
depending on ratios of proper scales, see \cite{BDFS2}.

\section{Conclusions}
 QCD approach to MPI leads to the notion of generalized double parton distributions, $_2$GPDs. 
 Higher order logarithmic QCD corrections to $_2$GPDs can be assembled via parton evolution equations derived in \cite{BDFS1,BDFS2} in the
 leading collinear approximation. Detailed formulae for total cross sections and differential distributions of four jet production 
 in the back-to-back kinematics can be found in \cite{BDFS2}.

 In order to reliably extract MPI contributions and get hold of parton correlations inside nucleon, one has to use a different strategy 
 from that developed and promoted by Tevatron experiments \cite{Tevatron1,Tevatron2,Tevatron3}. 
 Tevatron methods were based on measurement of {\em angular}\/ correlations between jet imbalance momenta proposed in \cite{Berger}. 
 Such characteristics, however, are sensitive to non-perturbative physics and are strongly affected by experimental efficiencies of jet reconstruction. 
 They are difficult (if at all possible) to control theoretically and should be replaced by studies of correlations in transverse momenta rather than angles.


\begin{footnotesize}

\end{footnotesize}

\end{document}